\documentclass{emulateapj}

\def \etal         {{et~al. }}
\def \h2         {\hbox{H$_2$}}

\def\approxlt{\lower.2em\hbox{$\buildrel < \over \sim$}}
\def\approxgt{\lower.2em\hbox{$\buildrel > \over \sim$}}

\def \lfir          {\hbox{$L_{\rm FIR}$}}

\newcommand{\lsun}{\mbox{L$_\odot$}}
\newcommand{\msun}{\mbox{M$_\odot$}}
\newcommand\cmv{\mbox{cm$^{-3}$}}
\newcommand{\ee}[1]{\mbox{${} \times 10^{#1}$}}
\newcommand{\eten}[1]{\mbox{$10^{#1}$}}

\newcommand{\tbn}{\tablenotemark}
\newcommand{\mean}[1]{\mbox{$\langle#1\rangle$}} 
\newcommand{\td}{\mbox{$T_D$}}

\newcommand{\msunyr}{\mbox{M$_\odot$ yr$^{-1}$}}
\newcommand{\ldust}{\mbox{$L_{dust}$}}
\newcommand{\mgas}{\mbox{$M_{gas}$}}
\newcommand{\nd}{\nodata}
\newcommand{\snu}{\mbox{$S_{\nu}$}}

\begin{document}

\title {Continuum Observations at 350 Microns of High-Redshift Molecular
Emission Line Galaxies}

\author {Jingwen Wu\altaffilmark{1,}\altaffilmark{2},  
Paul A. Vanden Bout\altaffilmark{3}, 
Neal J. Evans II\altaffilmark{1},  
Michael M. Dunham\altaffilmark{1}}

\altaffiltext{1}{Department of Astronomy, University of Texas, Austin TX 78731,
USA; nje@astro.as.utexas.edu, mdunham@astro.as.utexas.edu}
\altaffiltext{2}{Harvard-Smithsonian Center for Astrophysics, 60 Garden St. 
MS78, Cambridge, MA, 02138, USA; jwu@cfa.harvard.edu}
\altaffiltext{3}{National Radio Astronomy Observatory,
Charlottesville, VA 22903, USA; pvandenb@nrao.edu.}

\begin{abstract}

We report observations of 15 high redshift ($z = 1-5$) galaxies at 350$\mu$m
using the Caltech Submillimeter Observatory and SHARC-II array detector.  
Emission was detected from eight galaxies, for which far-infrared luminosities,
star formation rates, total dust masses, and minimum source size estimates are
derived.  These galaxies have star formation rates and star formation
efficiencies comparable
to other high redshift molecular emission line galaxies.  The results are used
to test the idea that star formation in these galaxies occurs in a large number
of basic units, the units being similar to star-forming clumps in the Milky
Way. The luminosity of these extreme galaxies can be reproduced in a simple
model with (0.9-30)\ee6 dense clumps, each with a luminosity of 5\ee5 \lsun,
the mean value for such clumps in the Milky Way. 
Radiative transfer models of such clumps can provide reasonable matches to
the overall SEDs of the galaxies. They indicate that the individual clumps
are quite opaque in the far-infrared. Luminosity to mass ratios vary over
two orders of magnitude, correlating strongly with the dust temperature
derived from simple fits to the SED. The gas masses derived from the dust
modeling are in remarkable agreement with those from CO luminosities, 
suggesting that the assumptions going into both calculations are reasonable.

\end{abstract}


\section{Introduction} \label{intro}

Studies of star formation in the early Universe are important to an
understanding of galaxy formation and evolution.  Observations of quasar host
galaxies and sub-millimeter galaxies with detectable molecular line emission
offer an opportunity to extend such studies to high redshift ($z>2$), albeit
with a limited sample \citep[see][]{annrev05}.  Following that review,
we refer to these galaxies collectively as EMGs (Early Universe Molecular
Emission Line Galaxies). Several kinds of models of these objects have been
proposed. Dust in quasar host galaxies could be heated by the radiation
from the central AGN (e.g., \citealt{granato96}, \citealt{andreani99}).
While it is hard to rule out such models, general arguments tend to favor
starbursts as the main power source (e.g., \citealt{blain02}).
In addition, detailed studies of well-known sources 
have indicated that star formation dominates the contribution from the 
black hole (\citealt{rowan-robinson2000}, \citealt{weiss03}, 
\citealt{chapman04}, \citealt{weiss07}).

There are also variations among models based on star formation.
\citet{err03} proposed a model in which the far-infrared emission
arises from cirrus (relatively diffuse dust) heated by ultraviolet photons
leaking out from star formation regions. However, most recent analysts have
focused on models in which the far-infrared radiation arises from dust that
is intimately associated with a burst of star formation (e.g., 
\citealt{narayanan09a}).
In the embedded starburst models, the total luminosity of the
dust continuum emission is a measure of the star formation rate, and
the luminosity of the molecular line emission or the dust emission
at long rest wavelengths measures the amount of material
available for star formation.  

Here we report observations of the dust
continuum emission at 350$\mu$m wavelength, obtained at the Caltech 
Submillimeter
Observatory\footnote{The Caltech Submillimeter Observatory is supported by
the NSF.}.  
This observed wavelength falls
roughly near the peak of the emission in the rest frame of the objects observed
and is, therefore, a desirable wavelength for observations to determine
far-infrared luminosities. It will also provide stronger constraints on the
characteristic temperature of the far-infrared radiation, which can 
distinguish between cirrus models and models of embedded star formation,
as the former predict cool ($\td < 30$ K) dust \citep{err03}. 

This work extends previous studies at 350$\mu$m of
high redshift molecular line emission galaxies \citep{benford99, weiss03,
beelen06, wang08sharc2, wang08cont}.  Throughout this paper we have assumed a
$\Lambda$CDM cosmology with $H_0 = 70$ km s$^{-1}$ Mpc$^{-1}$, $\Omega_m =
0.27$, and $\Omega_{\Lambda} = 0.73$ \citep{spergel07}.

\section{Observations} \label{obs}

Table 1 lists the objects we observed.  They are, with two exceptions, objects
previously unobserved at 350$\mu$m,  chosen for the strength of their CO
emission and the strength of their long-wavelength dust
continuum emission to maximize the success of detection.  Table 1
also gives the coordinates observed, source redshifts from CO observations, 
the dates of the observations, the averaged zenith angle during observation 
with atmospheric opacities and integration times.

Observations were conducted in several runs during 2003 through 2007, using
the Submillimeter High Angular Resolution Camera II (SHARC-II) at the 10.4 m
telescope of the Caltech Submillimeter Observatory (CSO) at Mauna Kea,
Hawaii \citep{dowell03}. SHARC-II is a background-limited camera utilizing 
a ``CCD-style'' bolometer array with 12
$\times$ 32 pixels. At 350 \micron, the beam size is 8\farcs5, 
with a 2\farcm59 $\times$ 0\farcm97 field of view. Since the
atmospheric transmission is very sensitive to the weather at the higher
frequencies at which SHARC-II operates, most of our integrations were made
at small zenith angles and 
under the best weather conditions at Mauna Kea, during which the opacity at
225 GHz ($\tau_{225 GHz}$) was less than 0.06 in the zenith, which corresponds
to an opacity of 1.5 at 350 \micron\ (see Table 1). 
For most of our observations, the Dish Surface Optimization System
[DSOS, \citet{leong05}] was used to correct the dish surface figure for
imperfections and gravitational deformations as the dish moves in elevation.

All the raw data were reduced with the Comprehensive Reduction Utility for
SHARC-II, CRUSH - version 1.40a9-2 (\citealt{kovacs06a}, \citealt{kovacs06b}).
We used the sweep
mode of SHARC-II to observe all our sources. In this mode the telescope
moves in a Lissajous pattern that keeps the central regions of the maps
fully sampled. It works best for sources whose sizes are less than or
comparable to the size of the array, but causes the edges to be much noisier
than the central regions, and can often result in noise at the edges that
looks like real emission.  To compensate for this, we used ``imagetool,'' 
part of the CRUSH package, to eliminate the regions of each map
that had a total exposure time less than 25\% of the maximum.  This
eliminates most, but not all, of the spurious edge emission.

Pointing was checked every 1-2 hours during each run. The primary pointing
sources were planets such as Mars, Uranus, and Neptune, and their moons, 
for example,
Callisto. If no planets were available, we used secondary objects such as
CRL618 and IRC$+10216$. After averaging over all the runs, the blind pointing
uncertainty is 2\farcs1 for azimuth and 3\farcs1 for zenith angle.
We corrected the pointing after each check, so these uncertainties
actually represent upper limits. When reducing raw data with CRUSH, we also
applied a pointing correction based on the
statistics of all available pointing data (to remove the static error) and
several pointings before and after observing scans (to remove the dynamic
error).  This technique improves the calculated flux densities of point-like 
sources and further reduces the pointing 
uncertainty\footnote{http://www.cso.caltech.edu/$\sim$sharc/}. 

To obtain the total flux densities of the sources in units of Jy, we have 
used Starlink's
``stats'' package to measure the signal from targets in a 20$\arcsec$
aperture (an aperture large enough to include all the emission from 
high-z galaxies), and we measured the signal from calibrators 
in the same aperture. We used planets as calibrators whenever possible,
but some secondary calibrators were used when planets were not available.
The Flux Conversion Factors (FCF) for 
an 20$\arcsec$ aperture (C$_{20}$) is defined to be the total flux density 
of a calibrator source in Jy divided by the signal in that 
aperture from the calibrator in instrument units.  Since CRUSH already
includes an atmospheric correction, based on a fit to all calibrator
observations during the night, the flux density within 
the 20$\arcsec$ aperture is then obtained by simply multiplying the measured 
signal in the instrument units by C$_{20}$. The statistics of C$_{20}$ for 
all the calibrators over all our runs indicates a calibration uncertainty 
of 20\%. Detailed studies have been carried out of
calibration uncertainties  including uncertainties in the flux density
of the calibrator, the airmass of the source, and differences in atmospheric 
opacity at different times throughout the night. These studies were based on 
surveys towards low-mass Galactic cores observed on the same runs as
the galaxy observations. The results (Wu et al. 2007, Dunham et al. in prep), 
are that
the 20\% calibration error dominates over the airmass and atmospheric 
opacity errors. Therefore we take the systematic error to be 20\%, which
we add in quadrature to the random noise uncertainties.  

\section{Results} \label{results}

\begin{figure}
\includegraphics[width=6.0in,angle=0]{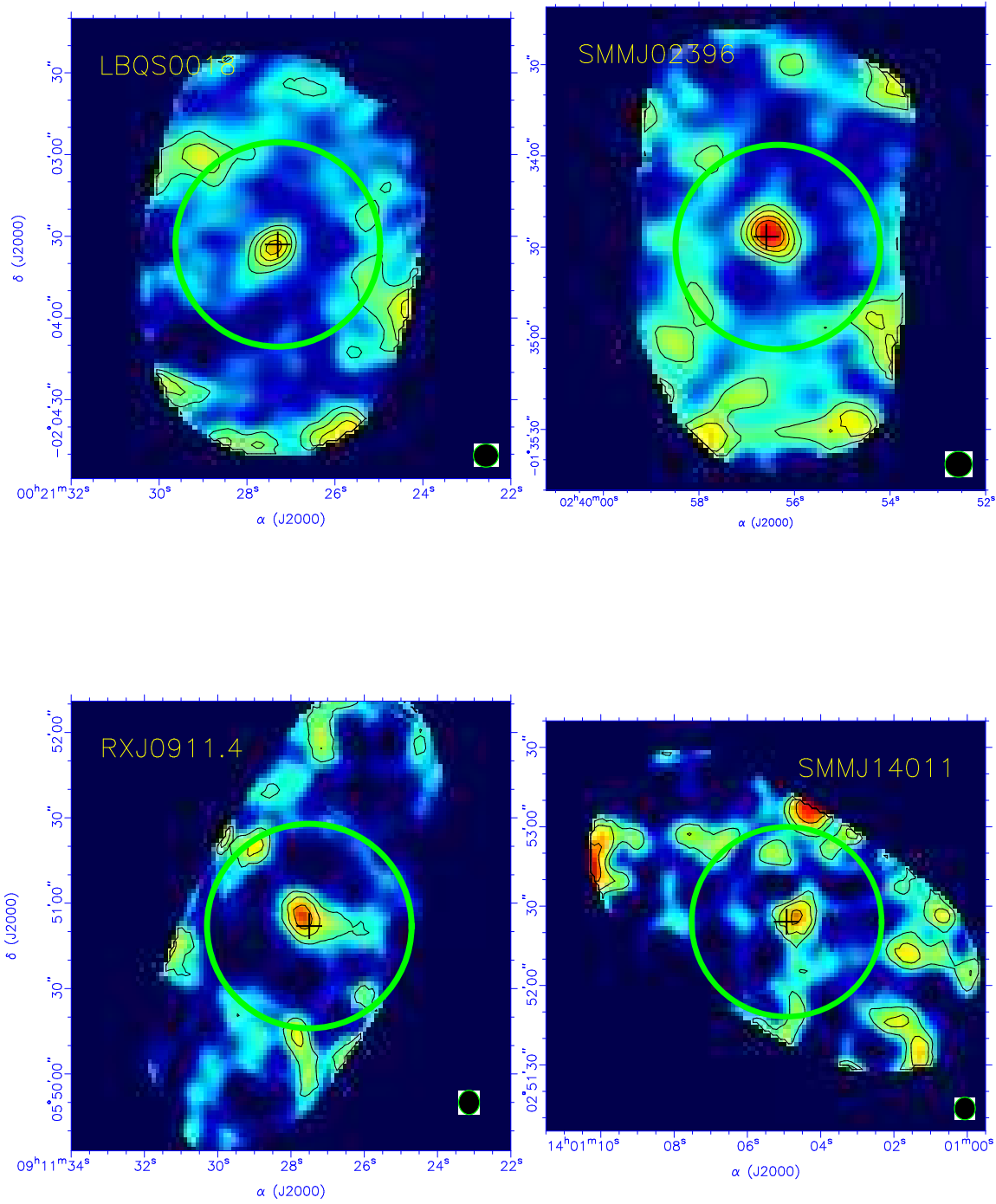}
\figcaption{\label{fig1}
Images of 350 \micron\ emission for four galaxies with detections at
$5 \sigma$ or better. The black circle at the lower right show the size of
beam. The colors of images from blue to red indicate increasing
intensity for the pixels. The intensity contours start from the $2 \sigma$
level, with a step of $2 \sigma$. The $1 \sigma$ levels for LBQS0018, SMMJ02396,
RXJ0911.4 and SMMJ14011 are 5, 6, 21, 10 mJy within a 20\arcsec\ aperture,
respectively. The black cross in each panel marks the position of the CO peak.
The green circles indicate the region within which detection is reliable.
Emission towards the edge is not reliable because of decreased
integration time and should be ignored.
}
\end{figure}

\begin{figure}
\includegraphics[width=6.0in,angle=0]{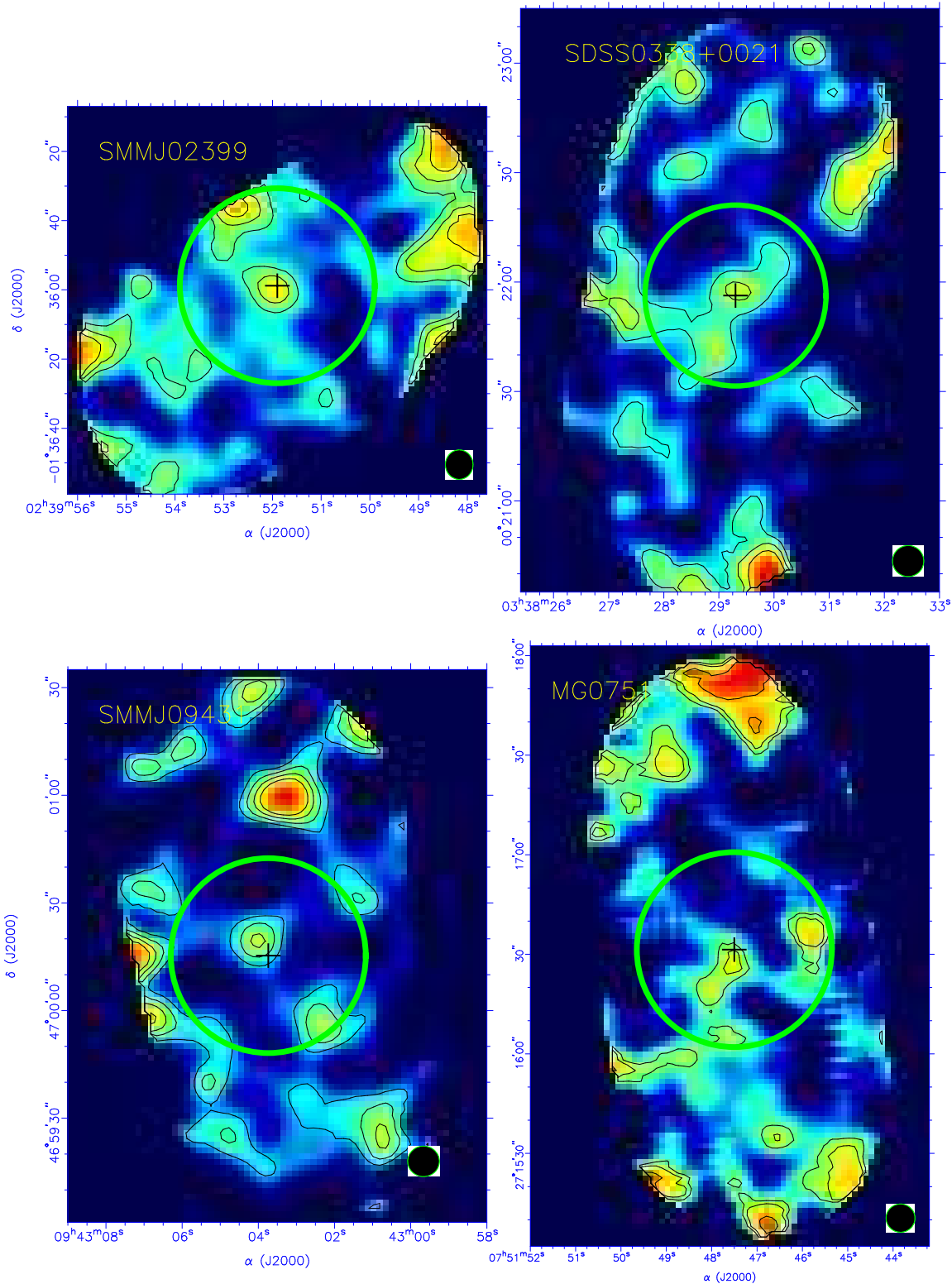}
\figcaption{\label{fig2}
Images of 350 \micron\ emission for four galaxies with detections at
less than $5 \sigma$. The black circles at the lower right show the size of
beam.  The colors of images from blue to red indicate increasing
intensity for the pixels. The intensity contours start from $1.5 \sigma$,
with a step of $1.5 \sigma$, except for MG0751, in which the
contour levels are $1.5 \sigma$ and $2.2 \sigma$. The $1 \sigma$ levels for
SMMJ02399, SDSS0038$+$0021, SMMJ09431 and MG0751 are 9, 9.5, 9, 16 mJy within
a 20\arcsec\ aperture, respectively. The black cross in each panel marks
the position of the CO peak. The green circles indicate the region within
which detection is reliable. Emission towards the edge is not reliable
because of decreased integration time and should be ignored.
}
\end{figure}

Of the 15 objects observed, we have four clear detections, 
($>5\sigma$), and four tentative detections ($2\sigma<\snu<5\sigma$). 
These are listed in Table 2 with the detected 
flux density
and noise in mJy, signal-to-noise ratio in $\sigma$, and detection
classification.  The 350$\mu$m images of the eight detected sources are given
in Figures \ref{fig1} and \ref{fig2}, where the circles indicate the 
region of reliable data reduction.

\begin{figure}
\includegraphics[width=5.5in,angle=0]{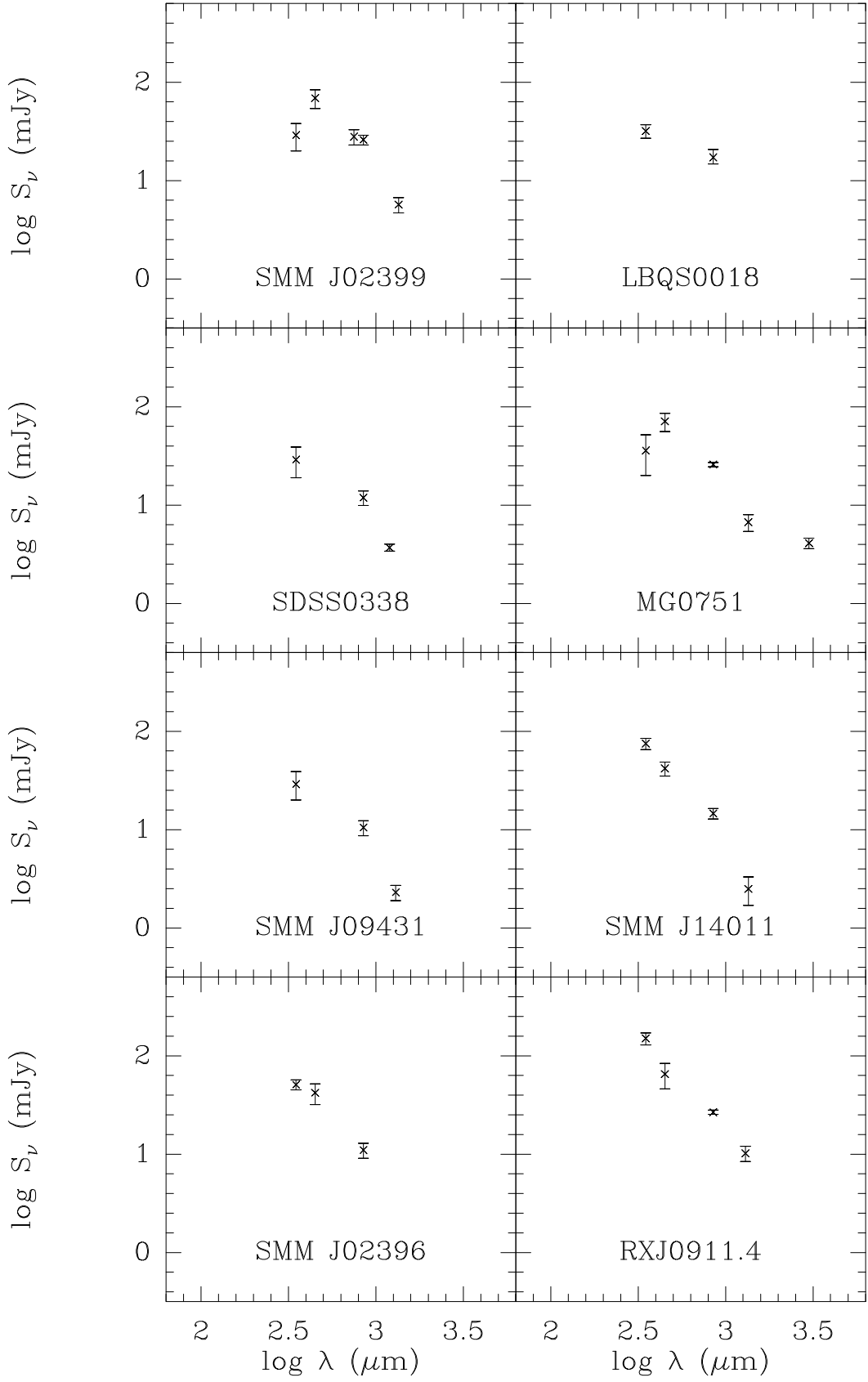}
\figcaption{\label{sedsobs}
The SEDs of all eight sources in the observed frame.
}
\end{figure}

We have collected data from the literature, focusing on long-wavelength
photometric data. These data will be used to model the sources, both with
simple graybody fits and with radiative transport models (\S \ref{models}).
The literature data are given in Table 3 and shown in Figure \ref{sedsobs}.

Table 4 gives properties derived from the fits to the SEDs, corrected
for magnification (noted as intrinsic properties) and redshift.
We fit optically thin graybody spectra to the 350--1200$\mu$m data points for
each source, using spectral energy distributions (SEDs)  of the form:
\begin{eqnarray}S(\nu_{rest}) & = & (1+z)^{-1}S(\nu_{obs}) \nonumber\\ & = &
S_0 \cdot  \left(\frac{\nu_{rest}}{1 THz}\right)^{3+\beta} \cdot
\frac{1}{exp(h\nu_{rest}/k\td) - 1}\end{eqnarray}
where $\beta$ is the emissivity
index, \td\ is the dust temperature in Kelvin, $S_0$ is an amplitude factor,
and
$S(\nu)$  is  in mJy.
In principle, it would be better to do the analysis without making
an assumption about the opacity of the dust.  However, the quality of the
350$\mu$m data is not sufficient to distinguish $\tau$ from $(1-e^{-\tau})$
in fitting SEDs to the data \citep [see] [] {kovacs06c}.  
Furthermore, the number of data
points is generally insufficient to fit more than two parameters; assuming
the dust is optically thin eliminates the source size from the fit.
We return to this issue in Sec. \ref{models}.

The luminosity of each source was calculated by integrating the best fit SED:
\begin{equation}
L = (2.5 \times 10^{-11})\ 4\pi D_L^2 \int S_\nu
\rm{d}\nu
\end{equation} 
where $D_L$ is the luminosity distance in Mpc and $L$ is the luminosity
in $L_{\sun}$.  In Table 4, the $L_{FIR}$ values are for integrals of the
SEDs over rest frequencies that correspond to observer frame wavelengths of
42.5--122.5 $\mu$m, using the definition of the FIR band given by
\citet{SandersMirabelrev}; the $L_{dust}$ values are for integrals over all
frequencies.  Since the shortest wavelength data we consider is $\lambda_{rest} = 58$
\micron, we assume that all the emission is coming from dust grains. In a few
cases, there may be contributions from other emission mechanisms to the longest wavelength data
(see \S \ref{models}), but these points generally do not drive the fit.
The dust temperature, $\td$, and $\beta$ are highly 
correlated and cannot be
separately determined with the few data points typically available for fitting
SEDs \citep[see][]{beelen06}.  In fitting SEDs to the observed data points, 
we held $\beta$ fixed, with the value $\beta = 1.5$.  This value 
is approximately that determined in data fits where there was sufficient 
information to vary $\beta$ \citep{beelen06}.  The value chosen does not 
have a dramatic effect on the calculated luminosity, which is determined 
largely by the observed 350 $\mu$m point.  In our data fits, changing the 
fixed value of $\beta$ by $\pm$ 20\% changed the luminosity by 
about  $+20/-50$\%.  Overall, we regard our calculated luminosities 
typically to be accurate to within a factor of 2--3.

Dust masses given in Table 4 were calculated from \begin{equation}M_d = (4.8
\times 10^{14})\  \frac{D_{L}^{2}}{(1+z)} \cdot \frac{S(\nu_{obs})}{
\kappa(\nu_{rest})B(\nu_{rest},T_D)}\end{equation} where $M_d$ is in
$M_{\sun}$, $B_{\nu}$ is the Planck function, and $\kappa_{\nu} =
\kappa_{0}(\nu/1 THz)^{\beta}$ is the dust mass absorption coefficient.  For
$T_D$ we used the results of the SED fit to the data.  
Similarly to calculated luminosities, the calculated values of
$M_d$ depend on the value of $\beta$ that is assumed.  For
$\nu_{rest}$ we have chosen  $\nu_{rest} = 0.35(1+z)$ THz, the rest frequency
for an observed wavelength of 850 $\mu$m, where the spectrum is solely dust
emission.  We set $\kappa_{0}$ = 1.3\  m$^{2}$ kg$^{-1}$ at 1 THz 
by interpolating in the table given by \citet{ossen94}.

Table 4 also lists the star formation rate (SFR) calculated from $L_{FIR}$
using $SFR(\msunyr) = (1.8 \times 10^{-10}) L_{FIR}(\lsun)$  
\citep {kennicuttrev},
$L_{dust}$/$M_{dust}$, and a minimum radius, $R_{min}^{350\mu m}$, for a source
of the dust radiation, assumed to be a disk seen face-on.  We calculated
$R_{min}^{350\mu m}$ by assuming the source to be optically thick
at $\nu _{rest} = 0.86(1+z)$ THz, the rest frequency for an observed
wavelength of 350\micron.
While some of the energy to heat the dust in the quasars may be supplied by 
accretion onto the black hole \citep{granato96}, we have attributed all 
of it to star formation. Consequently, the star formation rates we
derive could be overestimated.

\section{Discussion of Individual Sources} \label{sourcedisc}

\subsection{Detections}

\it LBQS0018 \rm is an optically identified quasar from the survey of
\citet{foltz89} detected in CO(3--2) with the IRAM Interferometer
\citep{izaak04}.  LBQS0018 is radio quiet.  

\it SMM J02396 \rm was detected in CO(3--2) emission by \citet{greve05} using
the IRAM Interferometer.   
The source is a ring galaxy identified by \citet{soucail99} from HST images.  
They measured the redshift of $z = 1.062$ and determined the magnification 
factor by the cluster Abell 370 to be $\mu = 2.5$.  
\citet{smail02} refer to the source as J02399-0134, rather than 
J02396-0134 as it has come to be labeled.  

\it RX J0911.4 \rm is a ROSAT source identified as a mini-broad absorption line
quasar by \citet{bade97}.  The quasar is strongly lensed ($\mu=22$) by a galaxy
at z = 0.8 \citep{kneib00}.  
RX0911.4 was detected in CO(3--2) at the Owens Valley Radio Observatory by
\citet{hainline04}.  The quasar is radio quiet.  

\it SMM J14011 \rm was the second SCUBA source to be detected in a molecular
emission line, namely CO(3--2) at the Owens Valley Radio Observatory
\citep{frayer99}.  It has been intensively studied since then in CO lines
\citep[see] [] {downes03}; the CI(1--0) line has been detected \citep{weiss05}.
The source is lensed, but the lens magnification
is very uncertain and model dependent, ranging from $\mu = 5 - 25$
\citep{downes03}.  

\subsection{Tentative Detections}

\it SMM J02399 \rm is a hyperluminous infrared galaxy \citep{ivison98}, the
first SCUBA source to be detected in molecular line emission, by
\citet{frayer99} at the Owens Valley Radio Observatory in CO(3--2).  From
higher angular resolution observations with the IRAM Interferometer,
\citet{genzel03} argued for a model with a very large ($r \geq 2$ kpc) disk, a
possibility that awaits still higher angular resolution confirmation.  
Our detection at 350 \micron\ of $29\pm9$ mJy is somewhat inconsistent
with the high value at 450 \micron\ \citep{ivison98}. (Fig. 3), 
even considering the large errors in each.

\it SDSS 0338 \rm is a high-redshift ($z = 5.03$) quasar discovered by
\citet{fan99}.  The CO(5--4) line was detected with the IRAM
Interferometer by \citet{maiolino07}.  The radio continuum has not been 
detected.  This source was previously detected at
350 \micron\ by \citet{wang08sharc2}, with  $S_{350\mu m} = 17.7 \pm 4.4$ mJy.
Our value of $29\pm10$ mJy is consistent within the uncertainties.  

\it MG0751 \rm is a quasar detected in CO(4--3) emission by
\citet{barvainis02a} using the IRAM Interferometer.  The quasar is highly
lensed, with magnification of 16.6 \citep{barvainis02b}.  
This is our weakest detection, with a S/N ratio of only 2.3, weaker than 
the 450 \micron\ detection.  

\it SMM J09431 \rm is a submillimeter galaxy detected in CO(4--3) emission by
\citet{neri03} using the IRAM Interferometer.  

\section{Testing the Idea of Basic Units} \label{units}

While high redshift starbursts are quantitatively more extreme than
anything in the local universe, some of their intensive properties, such as the
infrared luminosity per HCN line luminosity, are similar to those of
local starbursts and even cluster-forming dense clumps in our own
Galaxy  \citep{GaoSol04a, GaoSol04b, Wu05}.  This similarity has led to the
suggestion
that we can understand them in terms of a large number of basic units of
star formation, with these units being patterned on the well-studied
massive dense clumps in our Galaxy \citep{Wu05}. This model is similar
in spirit, though not in detail, to a model suggested by \citet{combes99}.

We use the data on this sample of EMGs to make a reality check on this
proposal. The idea is simple. Starting with some mean properties of
massive, dense clumps in our Galaxy, we divide the luminosity of the
EMG into $N_{clump}$ units and consider the consequences.

We take as our sample of Galactic massive, dense clumps the survey of
\citet{wu09}, which brings together data on 5 molecular lines
and infrared emission from over 50 regions, including some of the most
luminous regions in our Galaxy, such as W31, W43, W49, and W51. It is
important to include these, as the distribution of luminosities is strongly
skewed to lower values, and we seek a mean value. To be consistent,
\citet{wu09} calculated the luminosity of the massive clumps with the
same method as that used for high-z galaxies. The mean luminosity of
this sample is $(4.7\pm1.2)\ee5$ \lsun. \citet{Wu05} noted that
the ratio of infrared to line luminosity of these clumps matches that
of starburst galaxies only for clumps with a luminosity above about
$10^{4.5}$ \lsun. If we restrict the average to those clumps, the mean
value is $(6.3\pm1.6)\ee5$ \lsun. For simplicity, we will take
a value of $5\ee5$ \lsun\ as the standard value. Table 5 shows the
number of such average clumps that would be needed, calculated simply from
$N_{clump} = \lfir/5\ee5$. As explained in \S \ref{models}, we adopted
a value of $1\ee5$ \lsun\ for SMM J02396, with the corresponding increase
in the number of clumps.  Values of $N_{clump}$ range from 8.8\ee5 to 2.8\ee7.

Next, we ask if this large number of units would reasonably fit into
the space available. If not, the basic unit model could not work, and we
would have to turn to models with still denser dust, perhaps in a torus
around a black hole. We seek a minimum size estimate.
The most economical packing is spherical, but this
packing would likely result in high optical depth, even at submillimeter
wavelengths, so we also consider a two dimensional packing, in which
the units are placed in a disk, only one unit thick. The truth should
be somewhere between these two estimates. The minimum radii
are then given by $R_{sphere} = r_{clump}N_{clump}^{1/3}$ for the spherical
packing and $R_{disk} = r_{clump}N_{clump}^{1/2}$ for the disk packing.
Based on the sample in \citet{wu09}, $\mean{r_{clump}} = 1.13\pm1.09$
pc as measured by the size of the HCN 1-0 emission. For simplicity,
we take a standard value of 1.0 pc.
Sizes for spherical packing range from 95 to 300 pc, while those for
disk packing range from 940 to 5300 pc. These sizes are not unreasonable,
though the concept of a 5 kpc disk filled uniformly with massive dense
clumps may strain credulity. We emphasize that these are minimum size
estimates and that close packing of such clumps is very artificial. This
calculation is only a reality check and not a serious model. We
present such a model in the next section.

\section{Modeling the Units} \label{models}

The usual approach to modeling continuum emission from high-z starbursts
is to fit an optically thin, single temperature, modified black body.
The modification is
to assume an opacity that varies with wavelength as $\kappa_{\nu} =
\kappa_{\nu}(\lambda_{ref})(\lambda/\lambda_{ref})^{-\beta}$. This is the
procedure we have followed also, as described in Section \ref{results}. For
observers of Galactic star formation, this seems a quaintly anachronistic
procedure. Even a single region has a distribution of temperatures,
and opacities can be approximated by a power law only at long wavelengths.
In galaxies with large redshifts, the rest wavelengths observed in the
submillimeter are in the far-infrared, where dense clumps can be optically
thick.
These effects are commonly modeled by radiative transfer calculations that use
realistic opacities, place a luminosity source in the center of a
clump, and correctly account for optical depth effects.

By taking the basic unit idea a step further, we can explore the effects
of these complications by running radiative transport models for
an entity representing the mean clump. The observational properties
that constrain the models are, as usual, the flux densities. To perform
radiative transfer in the rest frame of the emitting galaxy, we scale all
wavelengths to the rest frame and we scale the flux densities according to

\begin{equation}
S_{\nu}(clump) = {5\ee5 \over \lfir} \cdot  \left({D_L \over 10 {\rm kpc}}
\right)^{2} \cdot  {1 \over \mu} \cdot S(\nu_{rest}),
\end{equation}
where $\mu$ is the lens magnification factor.

The choice of 10 kpc for the standard distance to the clump from the observer
is arbitrary and chosen just for convenience in comparison to clumps
in our Galaxy.

Because we keep the luminosity and size fixed at the values adopted above
($L = 5\ee5$ \lsun\ and $r_{clump} = 1.0$ pc), the only variables are those
that describe the mass and distribution of matter in the clump.
Models of continuum radiation from massive, dense clumps show that
that they are well modeled by power law density distributions
\citep [e.g.,] [] {mueller02, beuther02}:
\begin{equation}
n(r) = n_f(r)(r/r_f)^{-p}
\end{equation}
For example, \citet{mueller02}  fit radiative transfer models to data
for 31 sources and found $\mean{p} = 1.8\pm 0.4$, while \citet{beuther02}
found $\mean{p} = 1.6 \pm 0.5$ in the inner regions, steepening in
the outer regions. We can explore the effects of different choices
for $p$. We adopt single power laws over the whole clump. For connection
to previous work, we take $r_f = 1000$ AU. The mass of
the clump is then determined by the combination of $n_f$ and $p$.
The optical depth through the clump depends on the density distribution
and the assumed inner radius.
In modeling Galactic clumps, the longest wavelength point is generally
used to constain the mass because it is the most likely to be optically
thin. Thus, for a given value of $p$, $n_f$ is adjusted until the predicted
flux density matches the observations. For quasars, there could be 
contamination by free-free or synchrotron emission at the longest wavelength.
The only modeled source with a peculiar data point at long wavelengths is
MG0751 and we ignore that point in our fit.

We use the radiative transfer code by \citet{egan88}
as modified by us to produce outputs convenient to our purpose.
We use the opacities tabulated in column 5 of \citet{ossen94},
commonly referred to as OH5. The OH5 opacities have been shown to reproduce
well many observations of massive dense clumps. There is some evidence that they result in
underestimates of the mass by a factor of about 2, but this is still
uncertain and undoubtedly varies from region to region.

The luminosity in the models is represented by a single star at the center
of the mass distribution. The luminosity available for heating the dust
is set to the standard value of 5\ee5 \lsun. Depending on the effective
temperature of the star, its actual luminosity must be greater to account
for photons used to ionize an HII region. With no information to
constrain the nature of the forming star cluster, we simply set the
effective temperature to 44,000 K, roughly equivalent to an O5.5 star,
for which half the luminosity is in Lyman continuum photons. This choice
is largely irrelevant, as many studies have shown that the stellar photons
below the Lyman limit are rapidly absorbed and degraded to longer wavelengths,
so that the initial spectrum is rapidly erased 
\citep[e.g.,] [] {vdtak99}. 
The models also include heating from the outside by an interstellar
radiation field (ISRF). The ISRF is based on studies of the Milky Way
near us, as shown in \citet{evans01}. Since the radiation field is
likely to be much higher in these galaxies, we multiply the standard ISRF
by a factor of 100 as our default value.
We will explore the effect of changing this value.
The models also require an inner radius to the envelope. We adjust this value
so that the temperature there is in the range of 1000 to 2000 K, because
dust will evaporate at higher temperatures. The choice of inner radius has
no impact on the mass of the envelope or the flux density at long wavelengths, but
it has a major effect on the optical depth. The optical depth can strongly
affect the SED at shorter wavelengths, but the effect at the wavelengths we
are modeling is minor. The models result in values for $M_{gas}$, the clump mass, assuming
a gas to dust ratio of 100 by mass, the ratio
of the luminosity to mass, and the optical depth at all wavelengths, which
we characterize by $\tau_{100}$, the optical depth at 100 \micron.

\subsection{Example: Application to RXJ0911.4} \label{example}

As an example and to explore the effects of parameter choices, we consider
the case of RXJ0911.4, a highly lensed quasar at $z = 2.8$.
It has data available from 350 \micron\ to 3000 \micron, or from
92 to 790 \micron\ in the rest frame.

\begin{figure}
\includegraphics[width=6.5in,angle=0]{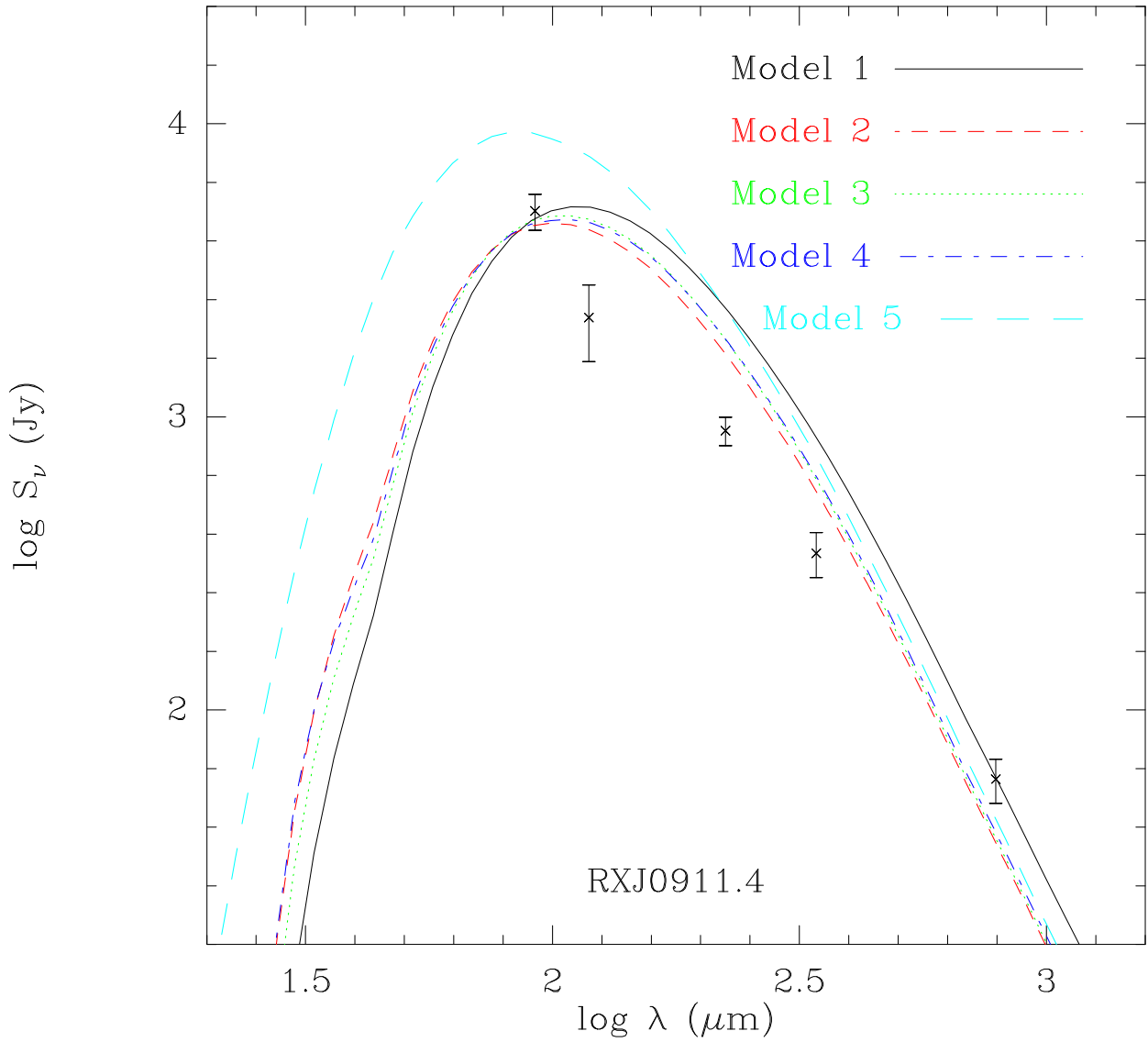}
\figcaption{\label{rxjmod}
The SED of the mean clump in RXJ0911.4 in the rest frame frequencies
and flux densities at a distance of 10 kpc. The models are described in the
text
and in Table 4.
}
\end{figure}

\begin{figure}
\includegraphics[width=6.5in,angle=0]{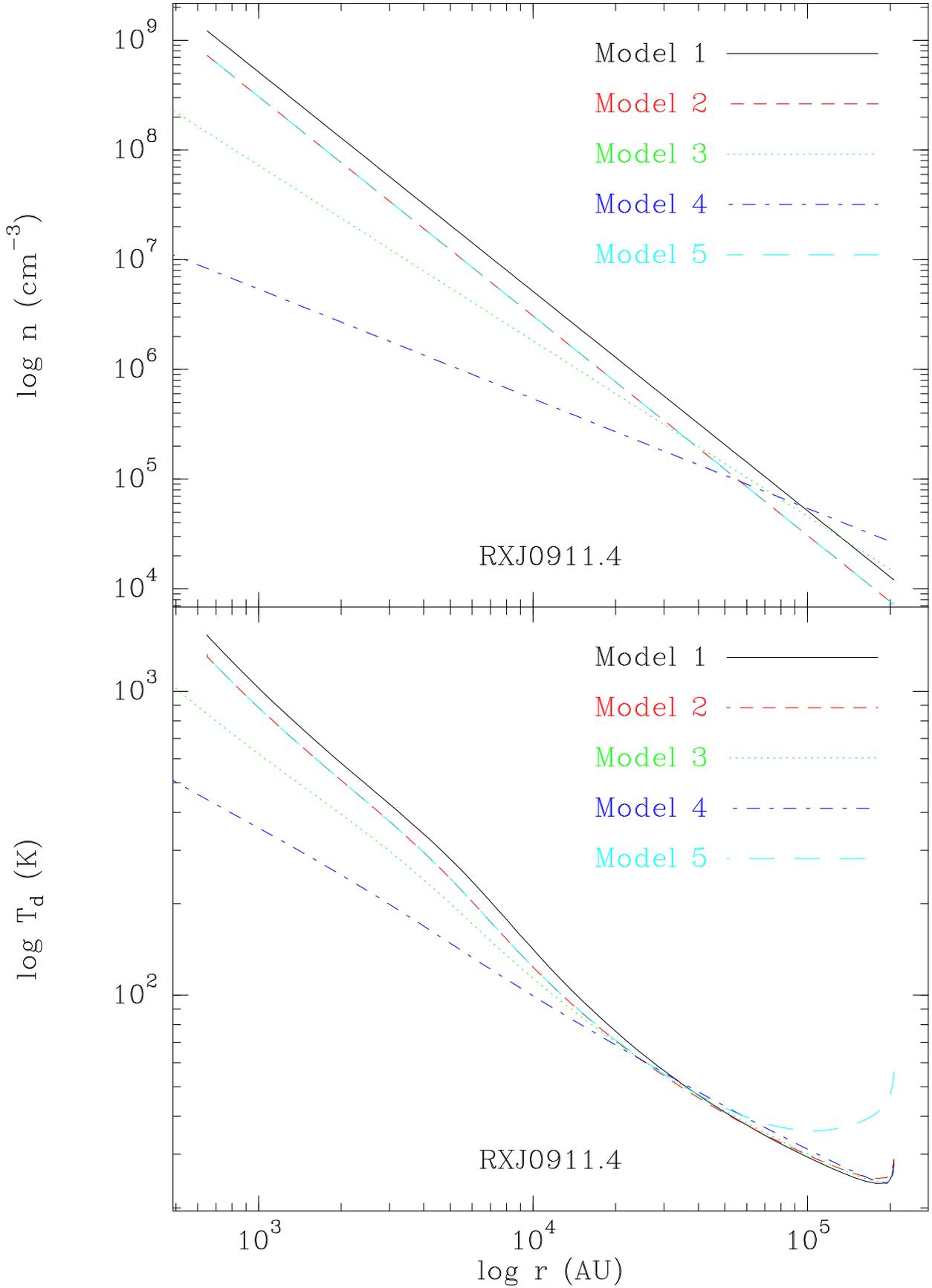}
\figcaption{\label{rxjdentemp}
The density and dust temperature are plotted versus radius for the
models of the mean clump in RXJ0911.4.
The models are described in the text
and in Table 4.
}
\end{figure}

Figure \ref{rxjmod}
 shows the SED with the data and several models for this source,
which are defined in Table 5.  Figure \ref{rxjdentemp} shows the run of 
density and dust temperature in each of the models.
In Model 1, we assume $p = 2$, and $n_f$ was adjusted to match the longest
wavelength point. It also matches the SHARC-II data, which corresponds
to a rest-frame wavelength of 92 \micron. However, it overestimates
all the data at intermediate wavelengths. If the data at rest-frame
790 \micron\ has a contribution from sources other than dust emission,
the mass could be overestimated. There was also a lower
value obtained at the same (observed) wavelength of 3000 \micron\
(\S \ref{sourcedisc}), which might suggest variability.
The rest of the models have a lower
value of $n_f$ and produce less emission at long wavelengths.
They provide a better compromise between the various observations.
In Model 3, we change $p$ to 1.6, and Model 4 has $p = 1.0$.
We adjust $n_f$ in each model to keep the predicted flux density about
the same at rest-frame 790 \micron, about 60\% of the (higher) observed value.
As is apparent, the value of $p$ has little effect on the SED at the
wavelengths with observations. However, the value of $p$ does affect
the optical depth. The values of optical depth at 100 \micron\ in the
rest frame are given in Table 5. These show that the usual assumption
of optically thin emission is highly dubious if the inner radius is as small
as our standard value. Data at shorter wavelengths are needed to
constrain the optical depth, though the assumption of spherical
symmetry then becomes an issue. For lower values of $p$, higher masses
are required to produce the same flux density at 790 \micron\ because the lower
values of optical depth result in more of the mass being at lower
temperatures. However, the difference in mass is about a factor of two,
as shown in Table 5.

We also show Model 5, which is the same as Model 2 except that we
increase the ISRF to 5000 times the local value. This produces an upturn
in the temperature at large radii, where the ISRF takes over, and the SED
is affected at the shorter wavelengths. In particular, the model now
is much too strong at the rest-frame wavelength of the SHARC-II observations.
The total luminosity from this model, based on integrating the SED, more
than doubles. Since this extra luminosity is coming from outside and would
represent luminosity from adjacent star forming regions or exposed luminous
stars, it is just a redistribution of the total luminosity of the galaxy. 
However, Figure
\ref{rxjmod} shows that a sufficiently well sampled SED could begin to 
distinguish these situations.

For the other sources, we restrict modeling to ISRF of 100 and $p = 2$.
We use the variation in $L_{dust}/M_{gas}$ found for the various values
of $p$ for RXJ0911.4 to estimate an uncertainty of 40\% on derived values
of $L_{dust}/M_{gas}$.

\subsection{Two Extreme Cases} \label{extremes}

Having established the basic dependencies on parameters for RXJ0911.4,
we discuss only a few salient features of models for the other sources.

\begin{figure}
\includegraphics[width=5.0in,angle=0]{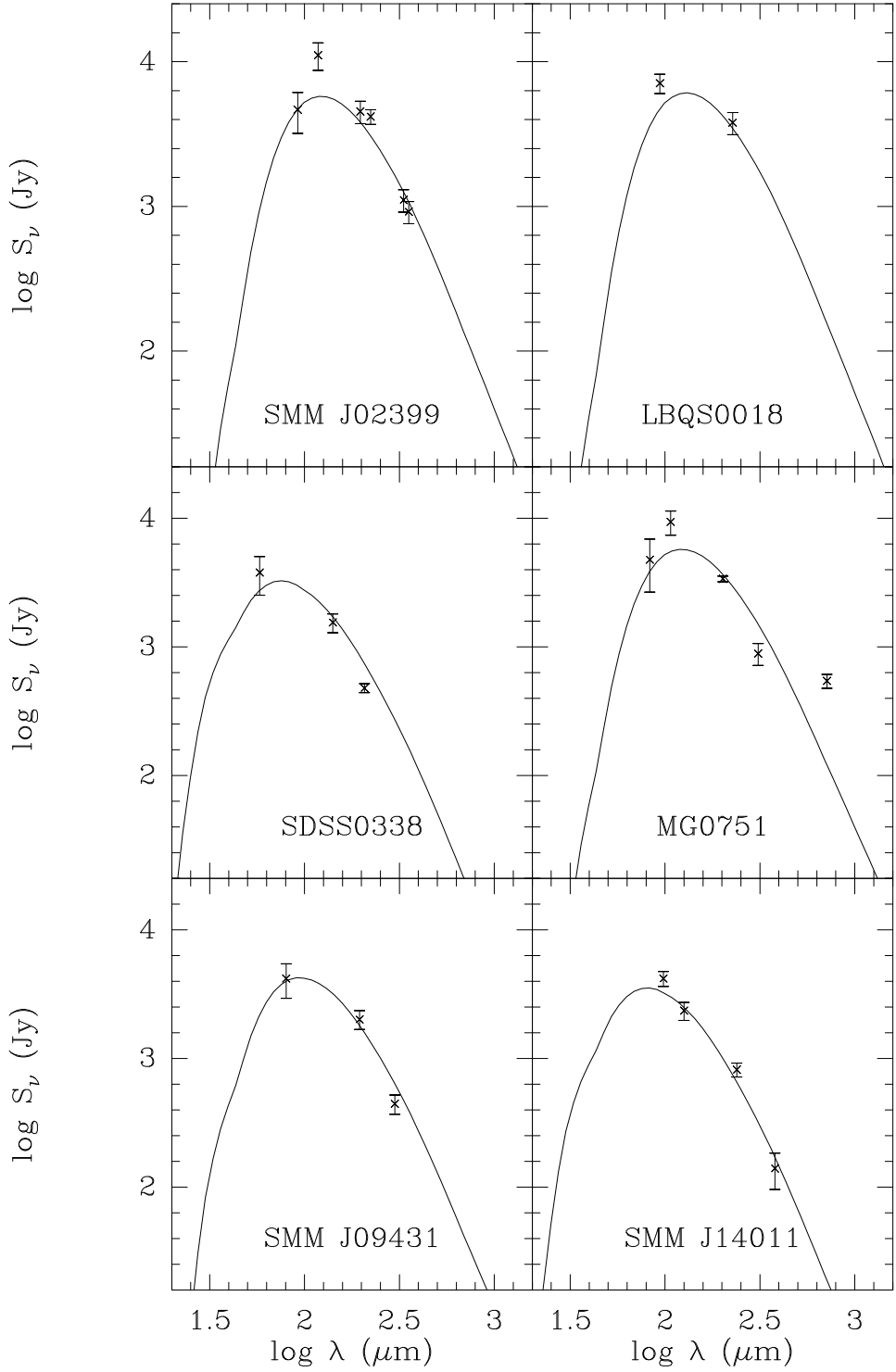}
\figcaption{\label{six}
The SEDs for the standard clump, placed at 10 kpc, in six galaxies 
with detections are plotted as points with errorbars. The flux densities
are shown at the rest wavelengths and the flux densities are scaled 
using equation 4.
The solid lines are the best-fitting models from the radiative transfer
modeling, as indicated in Table 4.
}
\end{figure}

SDSS0338, a quasar at $z = 5.0$, is the most extreme source in our sample.
With a $\lfir = 14\ee{12}$ \lsun\ and a $SFR = 2500$ \msunyr, it
exemplifies the most luminous sources known. As noted in Table 4,
it has the warmest SED, with $\td = 56$ K. The modeling of this source,
shown in Figure \ref{six}, is limited by having only 3 data points, none 
longer than
200 \micron\ in the rest frame. Data at longer wavelengths would be the
most valuable addition to the data base on this source.
Table 5 shows that the high value of
\td\ really implies a high luminosity to mass ratio, about five times that
of RXJ0911.4. With the usual interpretation of this ratio as a ``star formation
efficiency", a more useful way to look at the shape of the SED is in terms
of $\ldust/\mgas$ than in terms of a single temperature.

\begin{figure}
\includegraphics[width=6.5in,angle=0]{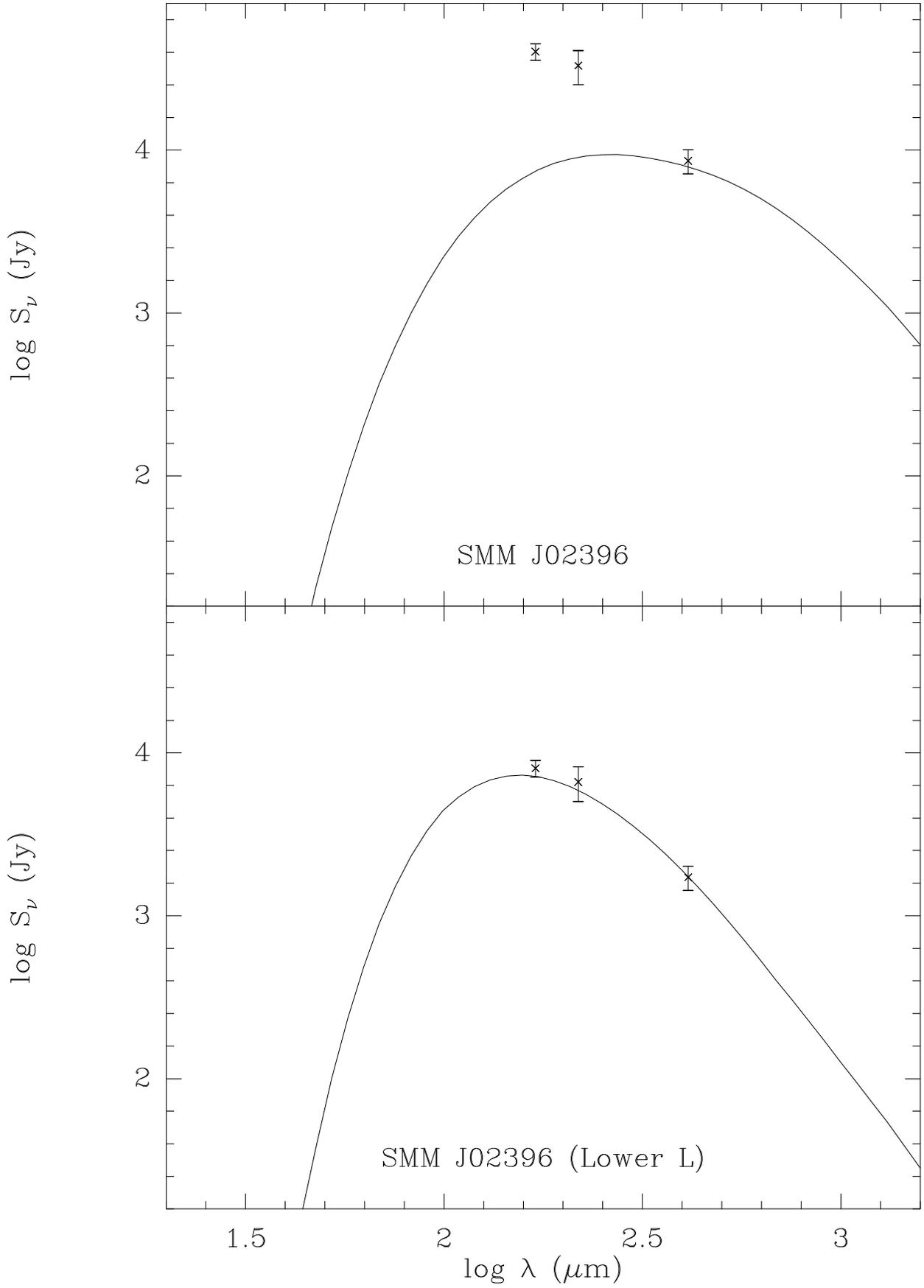}
\figcaption{\label{two}
(Top) The SED for the standard clump ($L = 5\ee5$ \lsun), placed at 10 kpc, 
in SMM J02396. No model was able to fit the flux densities.
(Bottom) The SED for a less luminous standard clump ($L = 1\ee5$ \lsun), 
placed at 10 kpc, in SMM J02396. Now a model can provide a reasonable fit.
}
\end{figure}

SMM J02396, at $z = 1.06$, is the coldest ($\td = 22$ K) and 
least luminous object in our sample,
with $\lfir = 2\ee{11}$ \lsun\ and a modest star formation rate,
$SFR = 37$ \msunyr, only about 10 times that of the Milky Way. Interestingly,
it {\it cannot} be fitted with the standard procedure. As shown in Figure
\ref{two}, the inferred flux densities for a standard clump are so high
that models fail to fit. To get even close to the longest wavelength point,
extremely high fiducial densities are needed, which cause huge optical depths.
The resulting cold SED indeed tries to reproduce the low value for \td\ from
the optically thin fitting, but it falls far below the shorter wavelength data.
Essentially, a source characterized by such a cold SED cannot have a luminosity
as high as 5\ee5 \lsun. The lower panel of Figure \ref{two} shows the result
of lowering the luminosity of the standard clump to 1\ee5 \lsun. Now the flux
densities are also five times lower (see equation 4) and can be fitted with
a reasonable model. The resulting ratio of luminosity to mass is very low.
In this galaxy, the mean clump would be less luminous and less star-forming
than the mean clump in our galaxy. The galaxy is detected at 350 \micron\ 
only because it has a huge amount of dense gas compared to the Milky Way.

This galaxy is a good candidate for the kind of cirrus model proposed by
\citet{err03}, where the dust responsible for the far-infrared emission
is far from the starburst. The fact that it is a ring galaxy \citep{soucail99}
is consistent with this kind of picture, but current observations
cannot tell whether the far-infrared emission is extended on the scale of
the 4\farcs8 diameter ring.

\section{Discussion} \label{discussion}

\subsection{Relations} \label{relations}

The value of $L_{dust}/M_{gas}$ can be related to a depletion time for
the dense gas and is often described as a star formation ``efficiency".
Values from the radiative transfer models range from 1.9 to 283,
with a mean of 105 and a median of 78. A similar calculation for dense
clumps in the Milky Way, as defined by HCN maps, yields similar values:
in a range of 3.3 to 398, the mean value is $117\pm20$, with a median of 97.
This similarity indicates that modeling extreme starbursts as an ensemble of
massive clumps similar to those in the Milky Way leads to a consistent
result.

\begin{figure}
\includegraphics[width=5.0in,angle=0]{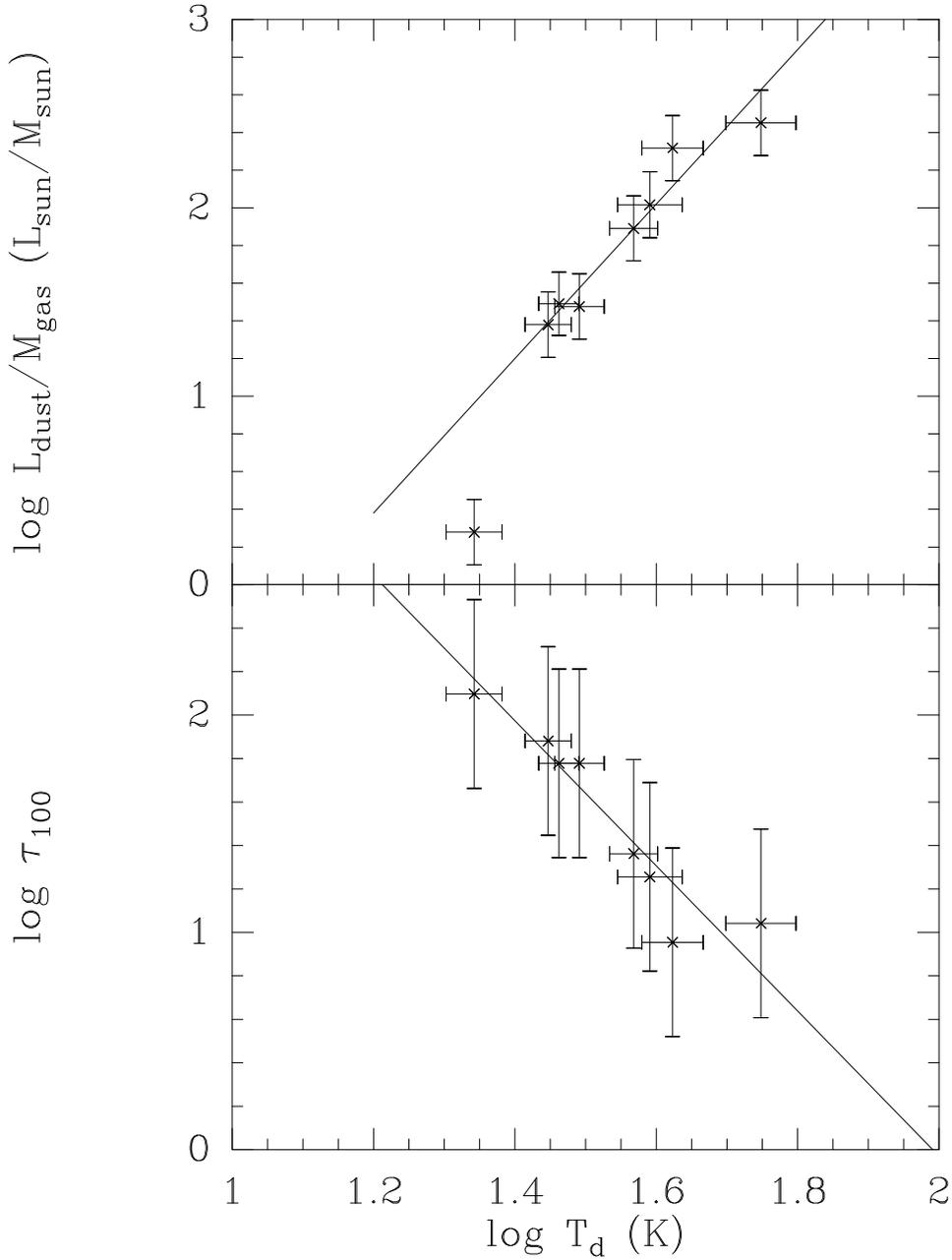}
\figcaption{\label{tdvslm}
(Top) 
The log of the luminosity to mass ratio from the radiative transport models is
plotted versus the log of the dust temperature derived from the isothermal, 
optically thin fit. The solid line is the least squares fit to the data given
by equation 6. 
The point at very low \td\ is SMM J02396, which has been
excluded from the fit.
(Bottom)
The log of the optical depth at 100 \micron\ from the best model is plotted
versus the log of the dust temperature derived from the isothermal,
optically thin fit. The solid line is the least squares fit to the data.
The uncertainties assigned to the optical depth were 100\%, to roughly 
indicate the dependence on model parameters.
The point at very low \td\ is SMM J02396, which has been
excluded from the fit.
}
\end{figure}

In the absence of detailed models, how should one interpret the dust
temperature derived from fitting the SED? 
There is a strong correlation between the fitted dust temperature and the
luminosity to mass ratio for the mean clump derived from the radiative
transport models (Figure \ref{tdvslm}). A least-squares fit in log space to the
points with uncertainties in both axes yields the following:
\begin{equation}
log (L_{dust}/M_{gas}) = (-4.55\pm1.54) + (4.11\pm0.97) log (\td)
\end{equation}
We have left SMM J02396 out of the fit, as it may represent a galaxy
dominated by cirrus emission. It is plotted in Figure \ref{tdvslm}
and clearly deviates from the fit to the other sources.
The value of $4.11\pm0.97$ can be understood in terms of the Stefan-Boltzmann
law, though we caution that $\td$ is, at best, an average over a large range of
real temperatures.
A similar relation is found for massive, dense clumps in the Milky Way
\citep{wu09}.

Using the relation from \citet{kennicuttrev} between star formation rate
and far-infrared luminosity, we can compute a depletion time for the dense
gas from

\begin{equation}
t_{dep}(y) = 5.6\ee{9}  (L_{dust}/M_{gas})^{-1} \\
 \approx 2.0\ee{14} \td^{-4.11}
\end{equation}
which ranges from 3.0\ee9 yr for SMM J02396 down to 2.0\ee7 yr for
SDSS0338. The mean value is 4.7\ee8 yr, or 1.1\ee8 yr, excluding the value
for SMM J02396.

The bottom plot in Figure \ref{tdvslm} shows the log of the optical
depth at 100 \micron\ from a single core versus the value of 
dust temperature from the simple fit to the SED. There is a strong
anti-correlation. (We have again left SMM J02396 out of the fit, but
it lies on the relation in this case.)
Galaxies with SEDs suggesting warmer dust can be
modeled with less opaque clumps as more luminosity is derived from
a smaller amount of mass.

\subsection{Comparison of Dust and CO} \label{compare}

The dust masses we have calculated from the simple, optically thin
isothermal fit (Table 4) assume an opacity given by models of
dust in dense clouds in the Milky Way \citep{ossen94}. The
gas masses emerging from the more detailed models and given in Table 5
make the same assumption about opacity and also assume a gas to dust
ratio of 100, as is usually assumed in the Milky Way. Both these
assumptions could in principle be quite wrong for high-z EMGs.
Masses computed from CO luminosities depend on a conversion to 
molecular hydrogen based on models of local ULIRGS, which is about a factor 
of 6 
times lower than for galaxies like the Milky Way \citep{annrev05}.
Thus substantial uncertainties attend these mass estimates and
consistency checks are worthwhile.

We have compared the masses derived from the simple fit to those
from the detailed radiative transport models by using a gas to dust
ratio of 100 to convert the former into gas masses. The mean ratio
of simple estimate to model mass is 0.64 with surprisingly small
dispersion. Thus, 
the simple fit gives a good estimate of the mass, even when the
assumptions are invalid. As long as there is a data point at wavelengths
long enough to be optically thin and in the Rayleigh-Jeans limit,
a simple fit provides an estimate that can agree with those from more
detailed models. If the models are correct, the masses from simple
fits should be scaled up by a mere 1.56.

Taking the gas mass estimates based on CO luminosity from \citet{annrev05},
we find that the mean ratio of the mass from CO to that from the models
is 0.60, with a bit more dispersion (minimum is 0.15 for SMM J02396
and maximum is 1.50 for SMM J14011). The agreement between these
two estimates is even more
surprising, as there are many reasons that they could differ.
The conversion from CO luminosity to mass is uncertain and
depends on internal cloud conditions like density and temperature
(e.g., \citealt{dickman86}, \citealt{solomon87}). The dust opacity
model could be different or the gas to dust ratio could differ.
For example, the two mass estimates could be reconciled if the gas to dust 
ratio is 167, which is 
probably within the range of uncertainty, even for the Milky Way
(see \citet{draine03} for discussion of abundance issues in grain models).
The general consistency despite all these possible differences
suggests that neither the standard conversion from CO
luminosity to gas mass nor the gas to dust ratio is likely to be far off.
The less satisfying 
possibility, that both conversions are wrong and the errors cancel out, 
cannot be ruled out.

Considering the possible differences in radiation environment, metallicity,
etc., the consistency between the masses from CO and from dust emission
is quite remarkable. We have used the submillimeter opacity values 
appropriate for coagulated, icy grains in dense molecular cores. 
The submillimeter opacity 
of the dust grains that characterize the diffuse interstellar medium in the
Milky Way is less by a factor of 4.8, which would result in masses from dust
emission
that are, on average, larger than those inferred from CO by a factor of 3. 
Assuming that the CO conversion factor is exactly correct, the dust 
properties in EMGs would appear to lie between those of the diffuse
ISM in the Milky Way and those of dense cores, but closer to the latter.

The remarkable thing about this comparison is that the masses from the models
include only quite dense gas (typically $n \geq \eten{4}$ \cmv, see Fig. 4), 
while the CO emission could easily arise in gas
of much lower density. The agreement suggests that essentially all 
the molecular gas in these EMGs is in dense clumps.

\subsection{Comparison to Other Models}

\citet{narayanan09a} have recently proposed a model for the formation of high
redshift submillimeter galaxies as the result of a starbursts triggered by
major mergers in massive halos. 
They modeled the time evolution of such mergers and 
found that the most luminous observed submillimeter galaxies could be modeled 
as 1:1 mergers in the most massive dark halos. As the two galaxies spiral
in toward the merger, SFRs of 100-200 \msunyr\ can be triggered. 
During final coalescence, they predicted a brief burst
(5\ee{7} yr) of very high SFR ($\sim 2000$ \msunyr). 
These predictions are broadly
consistent with our numbers for SFR (Table 4) and our mean depletion time of 
(1.1-4.7)\ee{8} yr. 
To make the most luminous submillimeter galaxies, they need to 
assume that all stellar clusters with ages less than 10 Myr are embedded 
in the material that is forming them, consistent with the assumptions of 
our modeling.

\citet{narayanan09b} have discussed the interpretation of CO lines from 
such galaxies.
They warn that the mass estimates from higher-J CO lines may underestimate 
the gas mass
by a factor up to 2. This would be consistent with our result above, but we 
believe that the sources of uncertainty discussed in \S 7.2 
are equally important. 

It is also interesting to compare our results to predictions of quite
different kinds of models. \citet{err03} modeled two of our sources
with their model of cirrus dominated far-infrared emission. In SMM J02399,
our data are consistent with their model, but in SMM J14011, their model
prediction is about a factor of 10 too low. This result is generally
consistent with the \td\ values in Table 5: SMM J02399 has
$\td = 29$ K, within the limit of 30 K that the dust achieves in their
models; SMM J14011 has $\td = 42$ K. On this basis, about half the sources
in our sample could be consistent, with SMM J02396 as a particularly
good case. Of course, models in which {\it some}
of the radiation arises in cirrus and some in dust surrounding the forming
stars could be relevant. Since the cirrus model is the logical extreme
of model 5 of RXJ0911.4, where we increased the ISRF, further constraints
on the SED could help to constrain such models.

While our data do not bear directly on models using radiation from
an AGN to heat a dusty torus, the general consistency of our models
with dust emission from star formation bolsters the case for the
embedded starburst models. 

\section{Summary} \label{summary}

We have observed 15 galaxies with redshifts from 1 to 5 at 350 \micron.
Four have detections at levels above $5 \sigma$, while four have
detections of lower significance. Far-infrared luminosities range from
2\ee{11} to 1.4\ee{13} \lsun, and inferred star formation rates range from
37 to 2500 \msunyr. From fits to optically thin, isothermal emission with
an opacity index, $\beta = 1.5$, characteristic dust temperatures range from
22 to 56 K. 

Because the dust is unlikely to be optically thin and isothermal, we consider
a picture in which the star forming regions are composed of a large number
($N_{clump} = 0.9 - 30 \times 10^6$)
of dense clumps, each with a luminosity equal to 5\ee5 \lsun, roughly the
mean value for massive star forming clumps in the Milky Way. A crude
calculation of the minimum size needed to pack such a large number of
clumps into a galaxy does not rule out such models.

Radiative transport models of standard clumps are then used to match
the observed SEDs, converted into the rest frame and scaled to a single
dense clump at a distance of 10 kpc, for ease of comparison to Milky Way
clumps. These models show that the individual clumps are likely to be
quite opaque at far-infrared wavelengths, but that the simple fits
do capture the total mass of emitting dust quite well.
The differences between sources lie primarily in the
ratio of luminosity to mass, which is commonly taken as a measure of star
formation ``efficiency" in extragalactic studies. Values of 2 to 283
are found for $L_{dust}/M_{gas}$. This ratio shows a strong correlation with
the value of \td\ from the isothermal, optically thin fit, resulting in
$L_{dust}/M_{gas} \propto \td^{4.11}$. The depletion timescales for dense gas
range from 3\ee9 y for SMM J02396 down to 2.0\ee7 y for SDSS0338.

\acknowledgements

We appreciate the support and assistance provided by the SHARC-II
team at Caltech during observation and data reduction: Darren Dowell,
Attila Kovacs, Colin Borys, and their colleagues. 
P.V.B. thanks the Department of Astronomy, University of Texas at Austin 
for hospitality shown during the course of this research.
This research was supported in part by NSF Grant AST-0607793 to
The University of Texas at Austin.

\begin{deluxetable}{llrccccc}
\tablenum{1}
\tablewidth{6.5in}
\tablecaption{Source List and Observations}
\tablehead{
\colhead{Source} & \colhead{R.A.} & \colhead{Decl.} & \colhead{Redshift} &
\colhead{Observation} &\colhead{Zenith}& \colhead{$\tau_{225 GHz}$} & \colhead{Integration}\\
\colhead{} & \colhead{(J2000)} & \colhead {(J2000)} & \colhead{(z)} &
\colhead{Date} & \colhead{Angle}& \colhead{(neper)} & \colhead{(hr)}
}
\startdata
LBQS0018 & 00:21:27.30 & -02:03:33.0 & 2.62 &11/05 & 20 &0.06 & 1.0  \\
& & & & 12/06 & 30 & 0.05 & 2.0  \nl
& & & & 10/07 & 25 & 0.06 & 2.0  \nl
SMM J02396 & 02:39:56.60 & -01:34:26.6 & 1.06 & 12/06 & 30 &0.04 & 2.0  \\
& & & & 10/07 & 25 &0.05 & 1.0  \\
SMM J02399 & 02:39:51.90 & -01:35:58.8 & 2.81 & 09/04 & 25 & 0.05 & 2.0  \\
SDSS0338 & 03:38:29.31 & 00:21:56.3 & 5.03 & 10/07 & 30 & 0.05 & 2.0  \\
MG0414 & 04:15:10.70 & 05:34:41.2 & 2.64 & 09/04 & 15 & 0.06 & 1.5  \\
4C60.07 & 05:12:54.75 & 60:30:50.9 & 3.79 & 09/03 & 45 & 0.06 & 1.5  \\
MG 0751 & 07:51:47.46 & 27:16:31.4 & 3.20 & 04/04 & 30 & 0.07 & 2.0  \\
& & & & 04/04 & 25 & 0.08 & 1.5  \\
RXJ0911.4 & 09:11:27.40 & 05:50:52.0 & 2.80 & 04/04 & 25 & 0.07 & 1.0  \\
& & & & 03/05 & 30 & 0.08 & 1.5  \\
SDSS0927 & 09:27:21.83 & 20:01:23.7 & 5.77 & 10/07 & 30 & 0.06 & 1.0  \\
SMM J09431 & 09:43:03.74 & 47:00:15.3 & 3.35 & 03/05 & 30 & 0.07 & 2.5  \\
& & & & 12/06 & 30 & 0.04 & 1.0  \\
& & & & 10/07 & 40 & 0.05 & 1.0  \\
BR0952 & 09:55:00.10 & -01:30:07.1 & 4.43 & 04/04 & 30 & 0.07 & 1.0  \\
& & & & 06/05 & 50 & 0.05 & 2.0  \\
SMM J14011 & 14:01:04.93 & 02:52:24.1 & 2.57 & 04/04 & 20 & 0.06 & 1.0  \\
& & & & 04/04 & 20 & 0.08 & 1.0  \\
SMM J16359 & 16:35:44.15 & 66:12:24.0 & 2.52 & 06/05 & 45 & 0.06 & 3.0  \\
6C19.08 & 19:08:23.30 & 72:20:10.4 & 3.53 & 09/03 & 55 & 0.06 & 3.0  \\
B3 J2330 & 23:30:24.80 & 39:27:12.2 & 3.09 & 09/04 & 25 & 0.06 & 1.0  \\
& & & & 06/05 & 50 & 0.05 & 1.0  \\

\enddata
\end{deluxetable}

\begin{deluxetable}{lcccl}
\tablenum{2}
\tablewidth{4.5in}
\tablecaption{Source Detections}
\tablehead{
\colhead{Source}  &
\colhead{Flux Density}  & \colhead{Noise} & \colhead{S/N}  \\
\colhead{} & \colhead{(mJy)} & \colhead{(mJy)} & \colhead{($\sigma$)} 
}

\startdata
LBQS0018 & 32 & 5 & 6.4   \\
SMM J02396 & 51 & 6 & 8.5  \\
SMM J02399 & 29 & 9 & 3.1   \\
SDSS0338 & 29 & 9.5 & 3.1  \\
MG0751 & 36 & 16 & 2.3  \\
RXJ0911.4 & 150 & 21 & 7.1  \\
SMM J09431 & 22 & 6.6 & 3.3   \\
SMM J14011 & 75 & 10 & 7.5  \\

\enddata
\end{deluxetable}

\begin{deluxetable}{lccccccccc}
\tabletypesize{\scriptsize}
\tablenum{3}
\tablewidth{6in}
\tablecaption{Measured Flux Densities of Detected Sources}
\tablehead{
\colhead{Source}  &  \colhead{350$\mu m$}$^{a}$    & \colhead{450$\mu m$} &
 \colhead{750$\mu m$} &    \colhead{850$\mu m$} &
 \colhead{1200$\mu m$} &   \colhead{1300$\mu m$} &
 \colhead{1350$\mu m$} &   \colhead{3000$\mu m$} & \colhead{Ref}$^{b}$
\\
\colhead{} & \colhead{(mJy)} & \colhead{(mJy)} &
\colhead{(mJy)} & \colhead{(mJy)} & \colhead{(mJy)} &
\colhead{(mJy)} & \colhead{(mJy)} & \colhead{(mJy)} & \colhead{}
}

\startdata
LBQS0018   & 32$\pm$5 &          &        &17.2$\pm$2.9&
&            &            &           & 1 \\
SMM J02396 & 51$\pm$6 & 42$\pm$10&        & 11$\pm$1.9 &
&            &            &           & 2 \\
SMM J02399 & 29$\pm$9 &69$\pm$15 &28$\pm$5& 26$\pm$3   &
&            &5.7$\pm$1.0 &           &  3 \\
SDSS0338   & 29$\pm$10&          &
&11.9$\pm$2.0&3.7$\pm$0.3&            &            &           & 1,4 \\
MG0751     &36$\pm$16 &71$\pm$15 &        &25.8$\pm$1.3&
&            & 6.7$\pm$1.3&4.1$\pm$0.5&5,6\\
RXJ0911.4  &150$\pm$21& 65$\pm$19&        &26.7$\pm$1.4&
&10.2$\pm$1.8&            &1.7$\pm$0.3& 6 \\
SMM J09431 &22$\pm$7  &          &        &10.5$\pm$1.8&           &
2.3$\pm$0.4&            &           & 7,8 \\
SMM J14011 &75$\pm$10 &41.9$\pm$6.9&      &14.6$\pm$1.8&
&            &2.5$\pm$0.8 &           & 9,10 \\
 
\enddata
\tablenotetext{a}{All the 350${\mu m}$ fluxes are from this work.}
\tablenotetext{b}{\footnotesize 1. \citet{priddey03a}, 2. \citet{smail02},
3. \citet{ivison98}, 4. \citet{carilli01}, 5. \citet{barvainis02a},
6. \citet{barvainis02b}, 7. \citet{cowie02}, 8. \citet{neri03},
9. \citet{ivison00}, 10. \citet{downes03}.}
\end{deluxetable}

\begin{deluxetable}{lcccccccc}
\tabletypesize{\scriptsize}
\tablenum{4}
\tablewidth{6in}
\tablecaption{Intrinsic\tbn{a} Derived Parameters: Dust Temperatures, Luminosities,
Masses; Star Formation Rates}
\tablehead{
\colhead{Source}  &  \colhead{Lens} & \colhead{$T_{dust}$}  &
\colhead{$L_{FIR}$} & \colhead{$L_{dust}$} & \colhead{$M_{dust}^{850\mu}$} &
\colhead{SFR} & \colhead{$L_{dust}$/$M_{dust}$} & \colhead{$R_{min}^{350\mu}$}
\\
\colhead{} & \colhead{Mag.} & \colhead{(K)} & \colhead{($10^{12}$\lsun)} &
\colhead{($10^{12}$ \lsun)} & \colhead{($10^8$ \msun)} &
\colhead{(\msun yr$^{-1}$)} & \colhead{($10^{4}$ \lsun/\msun)} & \colhead{(pc)}
}

\startdata
LBQS0018 & 1 & 28 & 2.9 & 4.5 & 9.6 & 520 & 0.53 & 1710 \\
SMM J02396 & 2.5 & 22 & 0.2 & 0.4 & 1.9 & 37 & 0.21 & 300 \\
SMM J02399 & 2.5 & 29 & 1.8 & 2.6 & 5.8 & 320 & 0.45 & 1030 \\
SDSS0338 & 1& 56 & 14 & 23 & 2.6  & 2500 & 8.9 & 775 \\
MG0751 & 17 & 31 & 0.4 & 0.6 & 0.9  & 72 & 0.75 & 470 \\
RXJ0911.4 & 22 & 37 & 0.8 & 1.0 & 0.4 & 140 & 2.5 & 435 \\
SMM J09431 & 1.2 & 39 & 4.3 & 5.5 & 3.1  & 770 & 1.8 & 830 \\
SMM J14011 & 5--25 & 42 & 2.2--0.4 & 2.9--0.6 & 0.7--0.15 & 400--90 & 4.1--40 &
445--200
 \\

\enddata
\tablenotetext{a}{Corrected for the effects of lensing.}
\end{deluxetable}

\begin{deluxetable}{lccccccccc}
\tabletypesize{\scriptsize}
\tablenum{5}
\tablewidth{6in}
\tablecaption{Model Parameters}
\tablehead{
\colhead{Source}  &  \colhead{$N_{clump}$} & \colhead{$R(sphere)$} &
\colhead{$R(disk)$} & \colhead{Model}  &
\colhead{$n_f$} & \colhead{$p$} & \colhead{$\tau_{100}$} &
\colhead{$M_{gas}$} & \colhead{$L_{dust}$/$M_{gas}$} \\
\colhead{} & \colhead{} & \colhead{(pc)} & \colhead{(pc)} &
\colhead{} & \colhead{(\cmv)}
& \colhead{} & \colhead{} &
\colhead{(\msun)} & \colhead{(\lsun/\msun)}
}

\startdata
LBQS0018  & 6.9\ee6 & 180 & 2410 & 4  & 1.3\ee9  & 2.0 & 76 & 2.1\ee4 & 24 \\
SMM J02396\tbn{a} & 2.0\ee6 &  125 &  1414 & 5 & 3.3\ee9 & 2.0 & 125  & 5.3\ee4 & 1.9 \\
SMM J02399    & 3.6\ee6 & 153 & 1910 & 2 & 1.0\ee9 & 2.0 & 60  & 1.6\ee4 & 31 \\
SDSS0338  & 2.8\ee7 & 302 & 5300 & 1 & 1.1\ee8   & 2.0 & 11  & 1.8\ee3 & 283 \\
MG0751    & 8.8\ee5 &  95 &  940 & 3 & 1.0\ee9   & 2.0 & 60  & 1.6\ee4 & 30  \\
RXJ0911.4  & 1.6\ee6 & 116 & 1260 & 1 & 6.7\ee8 & 2.0 & 39 & 1.1\ee4 & 47 \\
      \nd & \nd     & \nd & \nd  & 2\tbn{d} & 4.0\ee8 & 2.0 & 23 & 6.4\ee3 & 78 \\
    \nd   & \nd     & \nd & \nd  & 3 & 9.2\ee7 & 1.6 & 11 & 8.8\ee3 & 57 \\
    \nd   & \nd     & \nd & \nd  & 4 & 6.6\ee6 & 1.0 &  2 & 1.1\ee4 & 47 \\
    \nd   & \nd     & \nd & \nd  & 5 & 4.0\ee8 & 2.0 & 23 & 6.4\ee3 &
113\tbn{b} \\
SMM J09431    & 8.5\ee6 & 203 & 2920 & 2 & 3.0\ee8 & 2.0 & 18 & 4.8\ee3 & 104 \\
SMM J14011\tbn{c} & 4.4\ee6 & 162 & 2090 & 2 & 1.5\ee8 & 2.0 & 9  & 2.4\ee3 & 208 \\
\enddata
\tablenotetext{a}{This model uses a clump luminosity of 1\ee5 \lsun, five times
less thant the standard value, used for all other galaxies.}
\tablenotetext{b}{Model 5 is the same as Model 2, except that the ISRF has
been multiplied by 5000 instead of 100; the extra luminosity is all from
the external radiation.}
\tablenotetext{c}{Assumes a lens magnification of 5.}
\tablenotetext{d}{This model was used in correlations and statistics.}
\end{deluxetable}

\clearpage

\begin{thebibliography}{}

\bibitem[Andreani et al.(1999)]{andreani99} Andreani, P., 
Franceschini, A., \& Granato, G.\ 1999, \mnras, 306, 16

\bibitem[Bade \etal (1997)]{bade97} Bade, N., Siebert, J., Lopez, S., \etal
1997, \aap, 317, L13.

\bibitem[Barvainis et al.(2002a)]{barvainis02a}
Barvainis, R., Alloin, D., \& Bremer, M.\ 2002a, \aap, 385, 399.

\bibitem[Barvainis \& Ivison(2002b)]{barvainis02b}
Barvainis, R., \& Ivison, R.\ 2002b, \apj, 571, 712.

\bibitem[Beelen et al. (2006)]{beelen06}
Beelen, A., Cox, P., Benford, D.J. \etal 2006, \apj, 642, 694.

\bibitem[Benford et al.(1999)]{benford99}
Benford, D.J., Cox, P., Omont, A., \etal
1999, \apjl, 518, L65.

\bibitem[Beuther et al. (2002)]{beuther02}
Beuther, H., Schilke, P., Menten, \etal
2002, \apj, 566, 945.

\bibitem[Blain et al.(2002)]{blain02} Blain, A.~W., Smail, I., 
Ivison, R.~J., Kneib, J.-P., \& Frayer, D.~T.\ 2002, \physrep, 369, 111

\bibitem[Carilli et al.(2001)]{carilli01}
Carilli, C.~L., et al.\ 2001, \apj, 555, 625.

\bibitem[Chapman et al.(2004)]{chapman04} Chapman, S.~C., Smail, 
I., Blain, A.~W., \& Ivison, R.~J.\ 2004, \apj, 614, 671 

\bibitem[Combes et al.(1999)]{combes99} Combes, F., Maoli, R., \& 
Omont, A.\ 1999, \aap, 345, 369

\bibitem[Cowie et al.(2002)]{cowie02}
Cowie, L.~L., Barger, A.~J., \& Kneib, J.-P.\ 2002, \aj, 123, 2197.

\bibitem[Dickman et al.(1986)]{dickman86} Dickman, R.~L., Snell, 
R.~L., \& Schloerb, F.~P.\ 1986, \apj, 309, 326

\bibitem[Dowell et al.(2003)]{dowell03}
Dowell, C.D., Allen, C.A., Babu, R. S., \etal 2003, \procspie, 4855, 73.

\bibitem[Downes \& Solomon(2003)]{downes03}
Downes, D., \& Solomon, P.~M.\ 2003, \apj, 582, 37.

\bibitem[Draine(2003)]{draine03} Draine, B.~T.\ 2003, \araa, 41, 241

\bibitem[Efstathiou \& Rowan-Robinson(2003)]{err03} Efstathiou, A., \& 
Rowan-Robinson, M.\ 2003, \mnras, 343, 322

\bibitem[Egan et al. (1988)]{egan88}
Egan, M.P., Leung, C.~M.,
\& Spagna, G.F., Jr. 1988, Computer Physics Communications, 48, 271.

\bibitem[Evans et al. (2001)]{evans01}
Evans, N.J., II, Rawlings, J.~M.~C., Shirley, Y.~L., \etal
2001, \apj, 557, 193.

\bibitem[Fan et al.(1999)]{fan99}
Fan, X., et al.\ 1999, \aj, 118, 1.

\bibitem[Foltz et al.(1989)]{foltz89} Foltz, C.B., Chaffee,
F.H., Hewett, P.C., \etal
1989, \aj, 98, 1959.

\bibitem[Frayer et al.(1998)]{frayer98} Frayer, D.T., Ivison,
R.J., Scoville, N.Z., \etal
1998, \apjl, 506, L7.

\bibitem[Frayer et al.(1999)]{frayer99}
Frayer, D.T., Ivison N.J., Scoville, N.Z. \etal 1999, \apjl, 514, L13.

\bibitem[Gao and Solomon (2004a)]{GaoSol04a}
Gao, Y., \& Solomon, P.M. 2004a, \apj, 606, 271.

\bibitem[Gao and Solomon (2004b)]{GaoSol04b}
Gao, Y., \& Solomon, P.M. 2004b, \apjs, 152, 63.

\bibitem[Genzel et al.(2003)]{genzel03} Genzel, R., Baker,
A.~J., Tacconi, \etal
2003, \apj, 584, 633.

\bibitem[Granato et al.(1996)]{granato96} Granato, G.~L., Danese, 
L., \& Franceschini, A.\ 1996, \apjl, 460, L11

\bibitem[Greve et al.(2005)]{greve05}
Greve, T.~R., \etal 2005, \mnras, 359, 1165.

\bibitem[Hainline et al.(2004)]{hainline04} Hainline, L.~J., 
Scoville, N.~Z., Yun, M.~S., Hawkins, D.~W., Frayer, D.~T., 
\& Isaak, K.~G.\ 2004, \apj, 609, 61

\bibitem[Ivison et al.(1998)]{ivison98} Ivison, R.~J., Smail,
I., Le Borgne, \etal
1998, \mnras, 298, 583.

\bibitem[Ivison et al.(2000)]{ivison00} Ivison, R.J., Smail,
I., Barger, \etal
2000, \mnras, 315, 209.

\bibitem[Izaak (2004)]{izaak04} Izaak, K.G.  2004, private communication.

\bibitem[Kennicutt (1998)]{kennicuttrev}
Kennicutt, R.~C., Jr.\ 1998, \araa, 36, 189.

\bibitem[Kneib et al.(2000)]{kneib00} Kneib, J.-P., Cohen,
J.~G., \& Hjorth, J.\ 2000, \apjl, 544, L35.

\bibitem[Kov{\'{a}}cs (2006a)]{kovacs06a} Kov{\'{a}}cs, A. 2006, 
arXiv:0805.3928v2.

\bibitem[Kov{\'{a}}cs (2006b)]{kovacs06b}
Kov{\'{a}}cs, A. 2006b, PhD thesis, Caltech.

\bibitem[Kov{\'a}cs et al.(2006c)]{kovacs06c} Kov{\'a}cs, A., 
Chapman, S.~C., Dowell, \etal
2006c, \apj, 650, 592. 

\bibitem[Leong (2005)]{leong05}
Leong,M.M. 2005, URSI Conf. Sec., J3-J10,426

\bibitem[Maiolino et al.(2007)]{maiolino07}
Maiolino, R., et al.\ 2007, \aap, 472, L33.

\bibitem[Mueller et al. (2002)]{mueller02}
Mueller, K.E., Shirley, Y.L., Evans, N.J., II, \etal
2002, \apjs, 143, 469.

\bibitem[Narayanan et al.(2009a)]{narayanan09a} Narayanan, D., 
Hayward, C.~C., Cox, T.~J., Hernquist, L., Jonsson, P., Younger, J.~D., 
\& Groves, B.\ 2009, arXiv:0904.0004 

\bibitem[Narayanan et al.(2009b)]{narayanan09b} Narayanan, D., Cox, 
T.~J., Hayward, C., Younger, J.~D., \& Hernquist, L.\ 2009, arXiv:0905.2184

\bibitem[Neri et al.(2003)]{neri03}
Neri, R., et al.\ 2003, \apjl, 597, L113.

\bibitem[Ossenkopf and Henning (1994)]{ossen94}
Ossenkopf, V. \& Henning, T.  1994, \aap, 291, 943.

\bibitem[Priddey et al.(2003a)]{priddey03a} Priddey, R.S., Isaak,
K.G., McMahon, R.~G., \etal
2003a, \mnras, 339, 1183.

\bibitem[Priddey et al.(2003b)]{priddey03b}
Priddey, R.~S., Isaak, K.~G., McMahon, \etal
2003, \mnras, 344, L74.

\bibitem[Rowan-Robinson(2000)]{rowan-robinson2000} Rowan-Robinson, M.\ 
2000, \mnras, 316, 885

\bibitem[Sanders and Mirabel (1996)]{SandersMirabelrev}
Sanders, D.B., \& Mirabel, I.F. 1996, ARA\&A, 34, 749.

\bibitem[Smail et al.(2002)]{smail02} Smail, I., Ivison, R.J., Blain, A.~W.,
\etal
2002, \mnras, 331, 495.

\bibitem[Solomon \& Vanden Bout (2005)]{annrev05}
Solomon, P.M. \& Vanden Bout, P.A. 2005, ARA\&A, 43, 677.

\bibitem[Solomon et al.(1987)]{solomon87} Solomon, P.~M., Rivolo, 
A.~R., Barrett, J., \& Yahil, A.\ 1987, \apj, 319, 730

\bibitem[Soucail \etal (1999)]{soucail99}
Soucail, G., Kneib, J.P., Bezecourt, J., \etal 1999, A\&A, 343, L70.

\bibitem[Spergel et al.(2007)]{spergel07} Spergel, D.N., \etal 2007,
\apjs, 170, 377.

\bibitem[van der Tak et al.(1999)]{vdtak99} van der Tak,
F.~F.~S., van Dishoeck, E.~F., Evans, N.~J., II, \etal
1999, \apj, 522, 991.

\bibitem[Wang et al. (2008a)]{wang08sharc2}
Wang, R., Carilli, C.L., Wagg, J., \etal  2008a, \aj, 135, 1201.

\bibitem[Wang et al.(2008b)]{wang08cont}
Wang, R., Carilli, C.L., Wagg, J., \etal 2008b, \apj, 687, 848.

\bibitem[Wei{\ss} et al.(2003)]{weiss03}
Wei{\ss}, A., Henkel, C., Downes, D., \etal
2003, \aap, 409, L41.

\bibitem[Wei{\ss} \etal (2005)]{weiss05} Wei{\ss}, A., Downes, D., Henkel, C.,
\etal
2005, \aap, 429, L25.

\bibitem[Wei{\ss} et al.(2007)]{weiss07} 
Wei{\ss}, A., Downes, D., Neri, R., Walter, F., Henkel, C., Wilner, D.~J., Wagg, J., \& 
Wiklind, T.\ 2007, \aap,

\bibitem[Wu et al. (2005)]{Wu05}
Wu, J., Evans, N.J. II, Gao, Y., \etal
2005, ApJ, 635, L173.

\bibitem[Wu et al. (2007)]{Wu07}
Wu, J, Dunham, M. M.; Evans, N. J., II, Bourke, T. L., Young, C. H., 2007, \aj,
133, 1560

\bibitem[Wu et al.(2009)]{wu09}
Wu, J., Evans, N. J., II, Shirley, Y. L., \& Knez, C., in prep.

\end{thebibliography}
\end{document}